\documentclass[aps,prl,reprint,superscriptaddress,floatfix,nofootinbib,showkeys,showpacs,preprintnumbers]{revtex4-2}
\pdfoutput=1

\usepackage{float}
\usepackage{tabularx}
\usepackage{xspace}
\usepackage{listings}
\usepackage{lineno}
\usepackage{bm}
\usepackage{amsmath}
\usepackage{amssymb}
\usepackage[dvipsnames]{xcolor} 
\usepackage{graphicx}
\usepackage{comment}
\usepackage{natbib}
\usepackage{hyperref}
\usepackage{amsmath}
\usepackage{enumitem}
\setlist{wide, labelwidth=!, labelindent=0pt}
\usepackage{multirow}
\hypersetup{    
  colorlinks      = true,
  linkcolor       = {black},
  linkbordercolor = {white},
  citecolor       = {black},
  citebordercolor = {white},
  urlcolor        = {blue},
  urlbordercolor  = {white},
}

\usepackage{soul} 
\usepackage{amsmath}
\usepackage{amssymb}
\usepackage{xspace}
\usepackage{xifthen}
\usepackage{float}
\usepackage{mathtools}
\usepackage{relsize}
\usepackage{aas_macros}

\newcommand{\abacussummit}{\textsc{AbacusSummit}}

\makeatletter 
\newcommand{\LCASES}[1]{$\m@th\displaystyle{#1}$\hfil}
\newcommand{\CCASES}[1]{\hfil$\m@th\displaystyle{#1}$\hfil}
\newcommand{\RCASES}[1]{\hfil$\m@th\displaystyle{#1}$}
\makeatother

\newcases{ecases}{\quad}{\CCASES{##}}{\LCASES{##}}{\lbrace}{.}
\newcases{ecases*}{\quad}{\CCASES{##}}{{##}\hfil}{\lbrace}{.}

\graphicspath{{Figures/}}

\def\beq{\begin{eqnarray}}
\def\eeq{\end{eqnarray}}

\newcommand{\Omegam}{\Omega_{\rm m}}

\begin{document}

\preprint{APS/123-QED}

\title{What Galaxy Clusters Have to Say About Dynamical Dark Energy and $\mathbf{H_0}$}

\author{Andr\'{e}s N. Salcedo}
\email[]{ansalcedo@arizona.edu}
\affiliation{ Department of Astronomy/Steward Observatory, University of Arizona, 933 North Cherry Avenue, Tucson, AZ 85721, USA.}
\affiliation{Department of Physics, University of Arizona, 1118 East Fourth Street, Tucson, AZ 85721, USA.}
\author{Eduardo Rozo}
\affiliation{Department of Physics, University of Arizona, 1118 East Fourth Street, Tucson, AZ 85721, USA.}
\author{Hao-Yi Wu}
\affiliation{Department of Physics, Southern Methodist University, Dallas, TX 75205, USA.}
\author{Shulei Cao}
\affiliation{Department of Physics, Southern Methodist University, Dallas, TX 75205, USA.}
\author{Enrique Paillas}
\affiliation{ Department of Astronomy/Steward Observatory, University of Arizona, 933 North Cherry Avenue, Tucson, AZ 85721, USA.}
\affiliation{Instituto de Estudios Astrof\'isicos, Facultad de Ingenier\'ia y Ciencias, Universidad Diego Portales, Av. Ej\'ercito Libertador 441, Santiago, Chile}
\author{Hanyu Zhang}
\affiliation{Waterloo Centre for Astrophysics, University of Waterloo,\\200 University Ave W, Waterloo, ON N2L 3G1, Canada}
\affiliation{Department of Physics and Astronomy, University of Waterloo,\\200 University Ave W, Waterloo, ON N2L 3G1, Canada}
\author{Eli S. Rykoff}
\affiliation{Kavli Institute for Particle Astrophysics \& Cosmology,\\
P. O. Box 2450, Stanford University, Stanford, CA 94305, USA}
\affiliation{SLAC National Accelerator Laboratory, Menlo Park, CA 94025, USA}

\date{\today}

\begin{abstract}
We show that, in flat $\Lambda$CDM, low-redshift structure probes --- cluster abundances, 3$\times$2-point analyses, and full-shape clustering --- are mutually consistent, jointly delivering precise constraints on $\sigma_8$ and $\Omega_{\rm m}$ that agree with geometrical datasets (CMB+BAO+SN). In $w_0w_a$CDM, adding clusters to the geometry dataset reduces the evidence for evolving dark energy while relaxing the $H_0$ tension, suggesting a $\Lambda$CDM evolution of the late-time Universe and a sound horizon that differs from its standard value.
\end{abstract}

\maketitle

\noindent \textbf{\emph{Introduction.}} --- The first DESI BAO data release reported tentative evidence for dynamical dark energy \citep{DESIDR1_BAO_et_al_2025}, which was strengthened by DR2 \citep{DESIDR2_BAO_et_al_2025}. If confirmed, this signal would alter our understanding of the cosmic energy budget and its connection to fundamental physics. Here we examine what low-redshift structure-growth experiments imply for this DESI signal.

\begin{figure*}[!t]
\centering \includegraphics[width=1.0\textwidth]{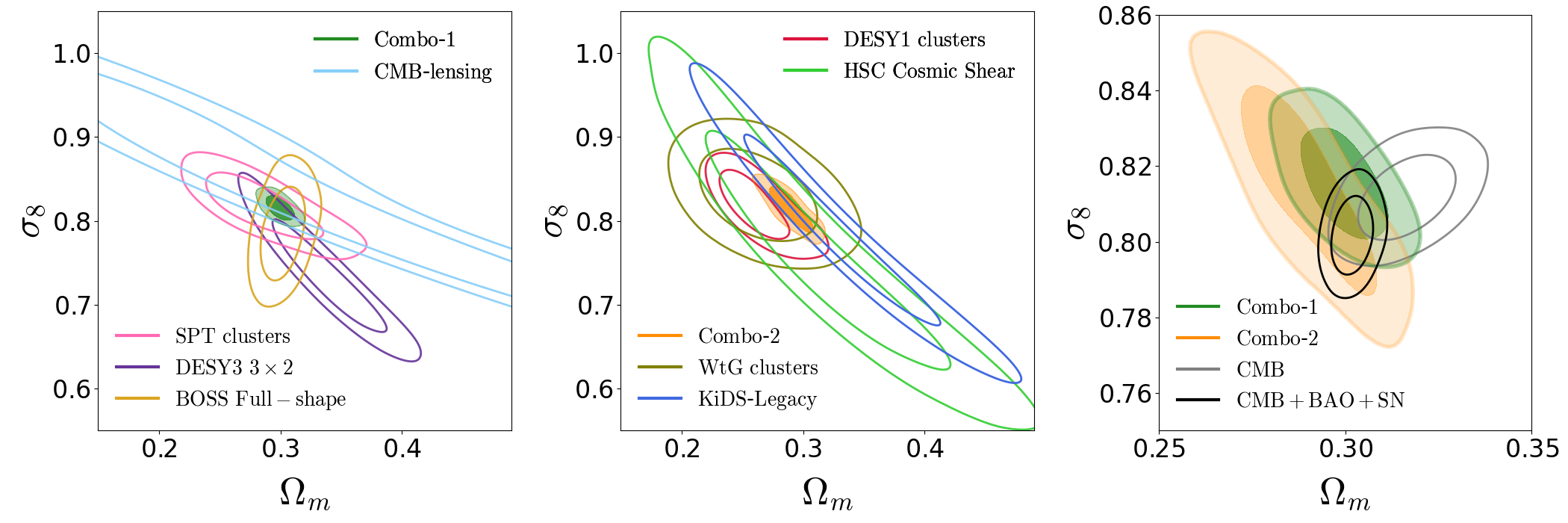}
    \caption{\textbf{Left and center:} Constraints from low-redshift experiments, as labeled in the legend. \textbf{Right:} Combined constraints from Combo-1 and Combo-2. The gray ellipse shows \textit{Planck} CMB constraints for flat $\Lambda$CDM, while the black contours add DESI BAO and DES Y5 SN to define the geometry-only data set.}
\label{fig:lowz}
\end{figure*}

\vspace{8pt}
\noindent \textbf{\emph{Data}} --- Our argument is based on cosmological posteriors from several analyses. The first eight are low-redshift structure-growth probes, all nearly independent, grouped into two sets of four denoted Combo-1 and Combo-2:
\begin{itemize}
    \item \textbf{Combo-1:} \vspace*{-7pt}
    \setlength{\itemsep}{-3pt}
    \begin{enumerate}
    \setlength{\itemsep}{-3pt}
    \item DES Y3 3$\times$2-point \citep{DES_3x2pt_2021}.
    \item SPT-SZ cluster abundances \citep{Bocquet_et_al_2024}. 
    \item SDSS full-shape galaxy clustering \citep{Kobayashi_et_al_2022}.
    \item CMB lensing \citep{Qu_SPA_lens_et_al_2025}. 
    \end{enumerate}

    \item \textbf{Combo-2:} \vspace*{-7pt}
    \begin{enumerate}
    \setcounter{enumi}{4}
    \setlength{\itemsep}{-3pt}
    \item DES Y1 optical galaxy clusters \citep{Salcedo_et_al_2025c}.
    \item RASS X-ray cluster abundances \citep{Mantz_et_al_2015}.
    \item HSC cosmic shear \citep{ZhangTQ_HSC3x2_et_al_2025}.
    \item KiDS-Legacy cosmic shear \citep{Wright_KIDS_et_al_2025}.
    \end{enumerate}
\end{itemize}
Details on each data set are given in the references above. Which experiment went into each combination was largely arbitrary, and our conclusions do not depend on this particular arrangement.  Additional discussion of our choice of experiments is provided in \cite{supp2}. We further complement Combo-1 and Combo-2 with the following data sets:
\begin{enumerate}
\setcounter{enumi}{8}
\setlength{\itemsep}{-3pt}
\item DESI BAO from DR2 \citep{DESIDR2_BAO_et_al_2025}. 
\item \textit{Planck} temperature and polarization data \citep{Planck_DR18_2020}.
\item DES Y5 supernovae (SN, $z>0.1$ only) \citep{DES5YSN_et_al_2024}.\footnote{A reanalysis of the DES Y5 SN data appeared on the arXiv as we were completing this work \citep{des25_sn_update}. The update reduces the evidence for dynamical dark energy, so adopting the new SN posteriors would likely strengthen our conclusions.}
\item Local $H_0$ measurements from the SH0ES \citep{shoes25} and CCHP \citep{cchp25} collaborations.
\end{enumerate}
We restrict the DES Y5 SN sample to $z>0.1$ because only these SNe were observed by DES; the low-redshift SNe come from historical, heterogeneous compilations. However, these low-$z$ SNe contribute to local $H_0$ estimates and are therefore included in item~12.

Throughout, CMB+BAO refers to the combination of DESI BAO and \textit{Planck} primaries (items 9 and 10), excluding CMB lensing (item 4 in Combo-1). We also define ``geometry-only'' below:
\begin{itemize}
    \setlength{\itemsep}{-3pt}
    \item \textit{Geometry-only}: CMB+BAO+DES Y5 SN.\footnote{This posterior was obtained using the public code {\sc{desilike}}: https://github.com/cosmodesi/desilike}
\end{itemize}

\vspace{8pt}
\noindent \textbf{\emph{Geometry~vs.~Growth in $\Lambda$CDM}}~---~
Fig.~\ref{fig:lowz} compares the cosmological posteriors from the four experiments in Combo-1 (left) and Combo-2 (middle). All four experiments from each combination are in good agreement. This is not by design: we can swap experiments between Combo-1 and Combo-2 without spoiling their internal consistency. Because the experiments in each combination are nearly independent, we combine their posteriors using {\sc CombineHarvesterFlow} \citep{Taylor_CombineHarvesterFlow_et_al_2024}.\footnote{For Combo-1, we use the posteriors from \citet{DES_SPT&3x2_et_al_2025} to combine the DES Y3 3$\times$2-point and SPT-SZ data sets.} Further details are given in \cite{supp2}.

The right panel of Fig.~\ref{fig:lowz} shows that Combos 1 and 2 agree with each other and are consistent with the geometry-only constraints. For reference, we also show the CMB-only constraint, illustrating that \textit{adding BAO and SN data to \textit{Planck} improves the consistency of geometry-only and structure-growth probes in $\Lambda$CDM.} This impressive agreement foreshadows a key question: how can we reconcile these results with the fact that $\Lambda$CDM is disfavored by both the Hubble tension and DESI's evidence for dynamical dark energy?

\vspace{8pt}
\noindent \textbf{\emph{Geometry vs.~Growth in $\mathbf{w_0}$$\mathbf{w_a}$CDM}} --- Most datasets in Combos 1 and 2 lack $w_0w_a$ runs, with the exception of DES Y1 clusters and DES 3$\times 2$-point.  Here, we focus on the geometry and clusters datasets. We also tested including DES 3$\times$2-point data, but found it had a negligible impact on the geometry+clusters posterior.

The simulation-based DES cluster analysis of \citet{Salcedo_et_al_2025c} relies on the \abacussummit\ simulations \citep{Maksimova_Summit_et_al_2021}, which span a limited set of cosmologies. When analyzing the clusters-only data we restrict the dark energy equation of state parameters to the \abacussummit\ range ($w_0\in[-1.271,-0.726]$ and $w_a\in[-0.628,0.621]$) to avoid emulator extrapolation. This restriction does not apply to the combined analysis.

\begin{figure*}
\centering \includegraphics[width=1.0\textwidth]{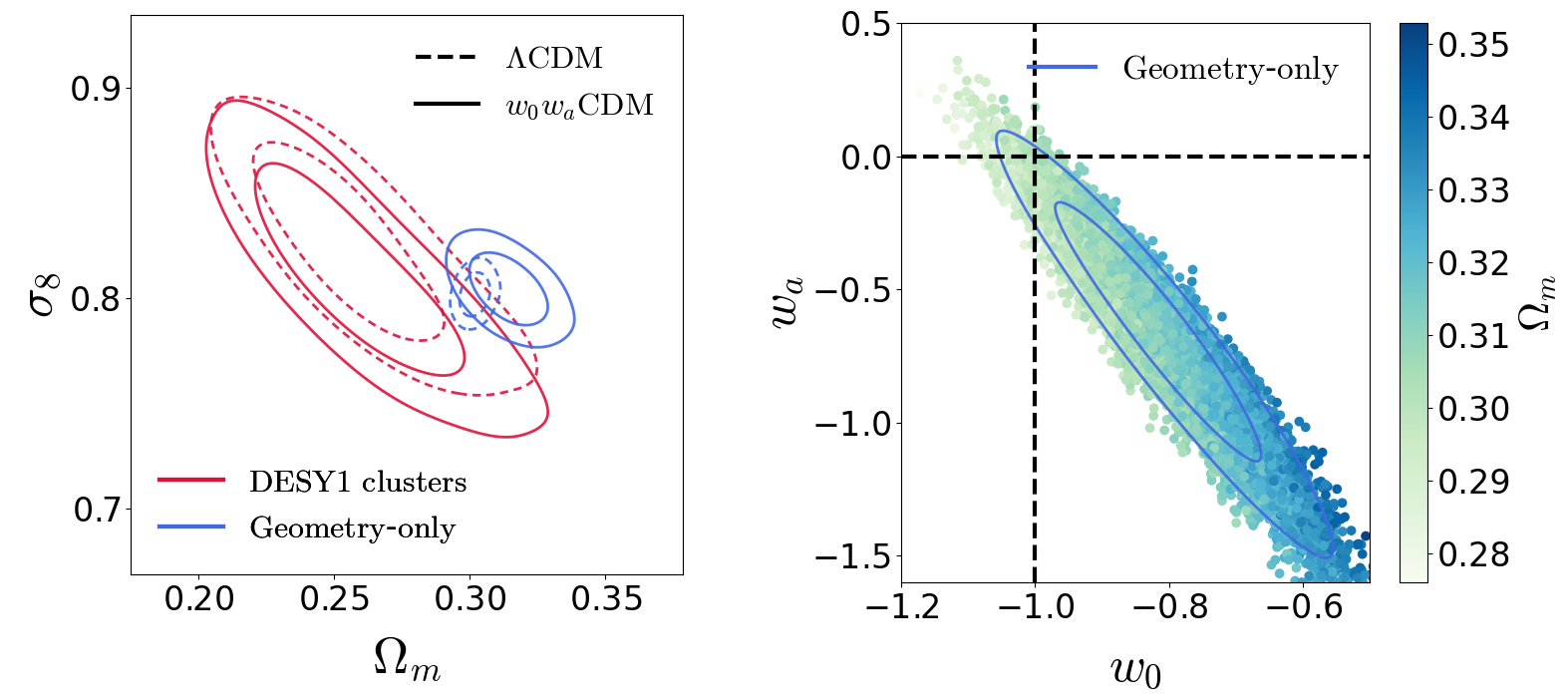}
    \caption{\textbf{Left:} Constraints from DES Y1 clusters and geometry in the $\sigma_8$–$\Omegam$ plane for $\Lambda$CDM and $w_0$–$w_a$ models. \textbf{Right:} 68\% and 95\% contours of the geometry-only posterior in $w_0$–$w_a$, with random samples color coded by $\Omegam$. See text for discussion.
    }
\label{fig:heuristic}
\end{figure*}

The 68\% and 95\% $\sigma_8$--$\Omegam$ contours for the DES-cluster and geometry-only posteriors are shown in the left panel of Fig.~\ref{fig:heuristic} for both $\Lambda$CDM and $w_0w_a$CDM. Since these are the two parameters that are well constrained by clusters, we evaluate the tension between the datasets in this subspace, finding $1.7\sigma$ and $1.9\sigma$ in $\Lambda$CDM and $w_0w_a$CDM, respectively \citep{Lemos_et_al_2021}. This justifies combining the two datasets.

\begin{figure*}[!t]
\centering 
\includegraphics[width=1.0\textwidth]{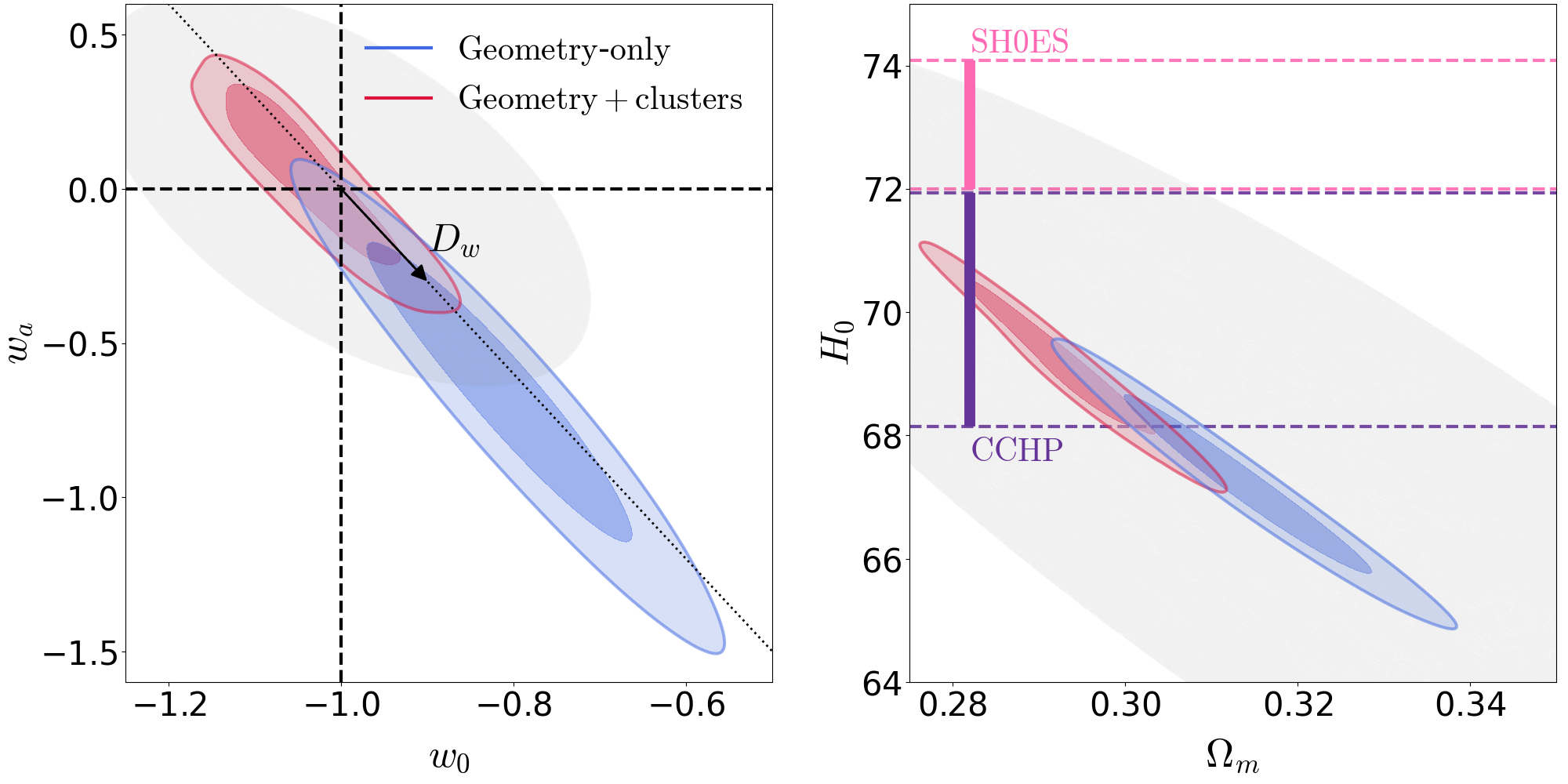}
    \caption{Posteriors for geometry-only and geometry+clusters in $w_0$–$w_a$ (left) and $\Omegam$–$h$ (right). Also shown are the $H_0$ constraints from SH0ES \citep{shoes25} and CCHP \citep{cchp25}. The left panel also shows the definition of $D_w$, the distance from $\Lambda$CDM along the line $3(1+w_0)+w_a=0$ (dashed black). The light gray ellipse shows the region where the clusters emulator is valid.}
\label{fig:w0wa_posteriors}
\end{figure*}

Fig.~\ref{fig:w0wa_posteriors} shows the geometry-only and geometry+DES-clusters posteriors in the $w_0$--$w_a$ (left) and $\Omegam$--$h$ (right) planes. The parameter space covered by the \abacussummit\ simulations is indicated by the light-gray ellipsoid, showing that the combined geometry+clusters constraints do not require emulator extrapolation. While geometry-only prefers an evolving dark energy, the geometry+clusters posterior peaks near $\Lambda$CDM.  That is, \textbf{clusters therefore dramatically reduce the preference for dynamical dark energy from the geometry-only data}.

The right panel of Fig.~\ref{fig:w0wa_posteriors} shows that adding DES clusters breaks the $\Omegam h^2$ degeneracy of the geometry-only data, yielding $\Omegam=0.293 \pm 0.007$ and $h=0.691 \pm 0.008$. This reduces the $H_0$ tension to $3.3\sigma$ and $0.5\sigma$ relative to SH0ES \citep{shoes25} and CCHP \citep{cchp25}. Conversely, \textbf{the dynamical dark energy models favored by the geometry-only data \emph{exacerbate} the Hubble tension.} This push toward $\Lambda$CDM from adding clusters or local $H_0$ measurements to the geometry-only results is puzzling given the strong $H_0$ tension in $\Lambda$CDM and the large $\Delta\chi^2$ preference for dynamical dark energy in DESI, an issue we return to in the discussion.

\vspace{8pt}
\noindent \textbf{\emph{Why Adding Clusters to Geometry Pulls the Posteriors Towards $\Lambda$CDM}} --- The left panel of Fig.~\ref{fig:heuristic} shows that clusters favor $\Omegam \lesssim 0.3$ irrespective of the dark-energy model. 
The right panel shows the geometry-only posterior in the $w_0$--$w_a$ plane with samples color-coded by $\Omegam$.  As one moves ``down and to the right,'' $\Omegam$ increases, reaching values excluded by the clusters posterior. 
Thus, clusters pull the geometry-only posterior to lower $\Omegam$, shifting $w_0$ and $w_a$ toward $\Lambda$CDM.

\begin{figure*}
\centering \includegraphics[width=1.0\textwidth]{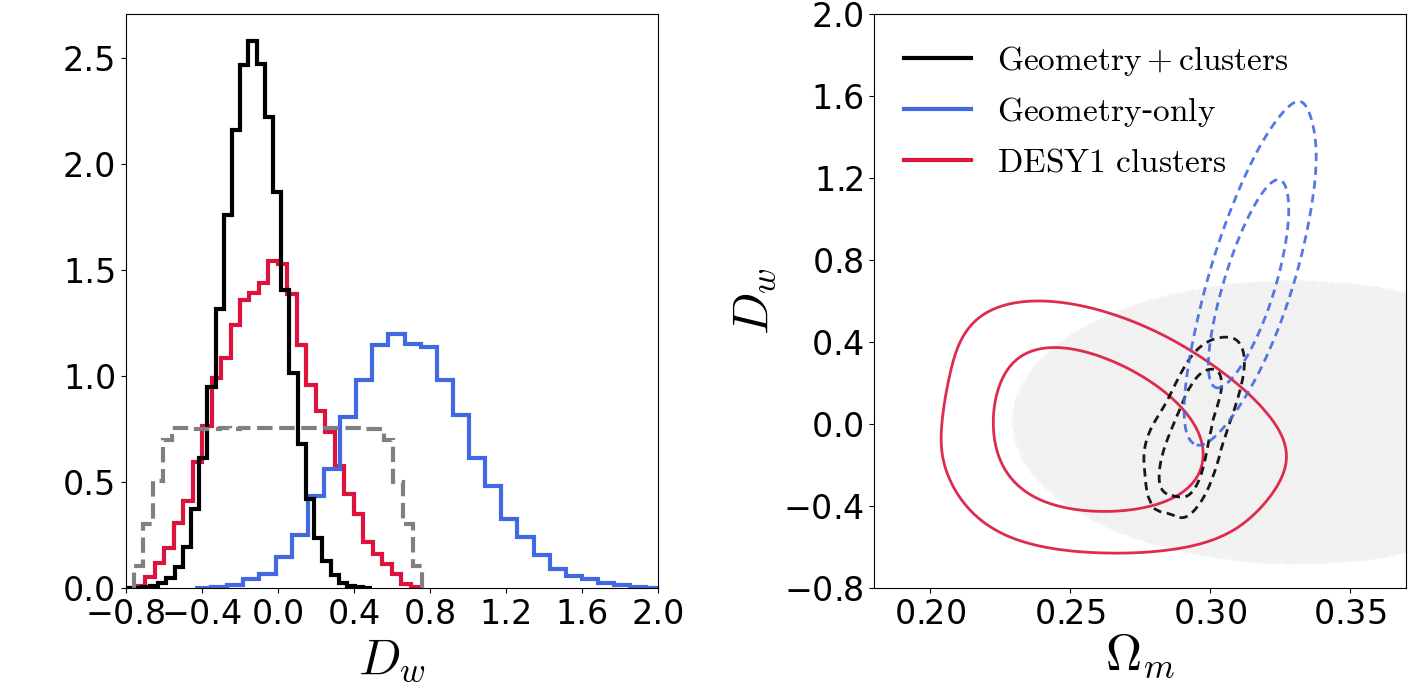}
    \caption{\textbf{Left:} Posterior distributions of $D_w$ for the DES clusters, geometry, and geometry+clusters data sets. Also shown is the prior in $D_w$ used to avoid emulator extrapolation in the clusters-only run (see text). The combined geometry+clusters chain uses a flat prior in all cosmological parameters. \textbf{Right:} The corresponding posterior distributions in the $D_w$-$\Omegam$ plane. The gray ellipse shows the parameter space over which the clusters emulator interpolates rather than extrapolates.}
\label{fig:Dw}
\end{figure*}

To further illustrate, we define $D_w$ as the distance from $(w_0,w_a)=(-1,0)$ along the degeneracy line $3(1+w_0)+w_a=0$ in $w$-space (see Fig.~\ref{fig:w0wa_posteriors}), with $D_w>0$ corresponding to ``down and to the right.'' The left panel of Fig.~\ref{fig:Dw} shows $D_w$ posteriors for DES clusters, geometry-only, and geometry+clusters, as well as the $w_0$ and $w_a$ prior used for the clusters-only run. The right panel shows the corresponding distributions in $D_w$--$\Omegam$. The geometry and clusters posteriors are nearly orthogonal, intersecting near the $\Lambda$CDM value $D_w=0$.

\vspace{8pt}
\noindent \textbf{\emph{Prospects for Confirmation}} --- We presented evidence against the dynamical dark energy interpretation of the DESI results. One way to strengthen these arguments is via independent estimates of the Hubble constant, e.g., from strong-lensing time delays \citep{tdcosmo25} or the energy-density method \citep{krolewski_etal25}. Robustly excluding low $h$ values ($h \lesssim 0.68$) cannot be reconciled with the evolving dark energy models favored by DESI without further modifications. Independent confirmation of low matter densities ($\Omegam \lesssim 0.3$) in $w_0w_a$CDM would also disfavor strongly evolving dark energy. Unfortunately, the DES 3$\times$2-point analysis does not yet sufficiently constrain $\Omegam$ in $w_0w_a$CDM to either confirm or refute our results.

Independent cluster surveys can also test our results. We assessed the feasibility of confirmation with SZ or X-ray selected clusters by removing the three lowest richness bins from our analysis. The reduced data vector had a similar (though weaker) impact on the $w_0$--$w_a$ posteriors, suggesting SZ/X-ray confirmation is possible.

As we were finalizing this manuscript, \citet{chudaykin_etal25} found that a full-shape analysis of the DESI DR1 data also favors lower matter densities, which shifts the $w_0$--$w_a$ posteriors back toward $\Lambda$CDM, in qualitative agreement with our findings. However, they report only a modest reduction in the significance of the deviation from $\Lambda$CDM.

\vspace{8pt}
\noindent \textbf{\emph{Summary and Discussion}} --- We showed that low-redshift structure data constrain $\Omegam \approx 0.29$ and $\sigma_8 \approx 0.81$ in $\Lambda$CDM (Fig.~\ref{fig:lowz}). These low-$z$ constraints are consistent with \textit{Planck}. Adding BAO+SN data to \textit{Planck} further improves the agreement between geometry and structure. This agreement is surprising in light of DESI's evidence for dynamical dark energy. Because most structure-growth analyses have not been performed in $w_0w_a$CDM, we extend our analysis to $w_0w_a$ by focusing on DES clusters. Adding DES 3$\times$2-point to the geometry+clusters combination has little impact on the posteriors.

Our posteriors for clusters-only, geometry-only, and geometry+clusters are summarized in Table~\ref{tab:constraints}. While geometry-only favors dynamical dark energy, geometry+clusters does not (Fig.~\ref{fig:w0wa_posteriors}, left). This shift arises because clusters favor lower matter densities than geometry-only, which pulls $w_0$ and $w_a$ toward $\Lambda$CDM. A similar trend was reported by \citet{chudaykin_etal25} using full-shape galaxy clustering in DESI DR1. The geometry+clusters posterior also alleviates the $H_0$ tension, recovering $h \approx 0.691 \pm 0.008$ in $w_0w_a$CDM.

The fact that a $w_0w_a$CDM fit yields consistent constraints from geometrical probes, galaxy clusters, and local estimates of the Hubble constant --- all while collapsing the posteriors around $\Lambda$CDM --- is striking. We interpret our results as supporting $\Lambda$CDM evolution of the Universe at late times. However, not all is well with the standard model: the $H_0$ tension in $\Lambda$CDM persists \citep[][]{h0book,divalentino21}, and the $\Delta\chi^2$ quantifying the DESI evidence for dynamical dark energy \citep{DESIDR2_BAO_et_al_2025} remains large. How can these observations be reconciled?

We see three possibilities. The first is that our clusters analysis is flawed and dynamical dark energy is rapidly evolving. However, the agreement among low-redshift structure probes argues against this, as does the fact that this solution exacerbates the $H_0$ tension. Ruling out this option requires independent measurements of $\Omegam$ from structure-growth probes in $w_0w_a$CDM \citep[e.g.][]{chudaykin_etal25}.

The second is that $\Lambda$CDM is incorrect and the true model deviates from $\Lambda$CDM just enough to bring $H_0$ into better, though still uncomfortable, agreement with local measurements. This solution moves the $w_0$--$w_a$ posterior ``up and to the left,'' likely exacerbating the DESI $\Delta\chi^2$ problem.

Lastly, late-time evolution could follow $\Lambda$CDM, implying that the standard-model sound horizon is incorrect, for example because of Early Dark Energy \citep{poulin_etal18} or modified recombination \citep{lynch_etal24}. Such models can also alleviate the negative-neutrino-mass problem \citep{negative_neutrinos,lynch_etal24}. Intriguingly, \citet{escudero_etal25} treated the sound horizon at the drag epoch as a free parameter, obtaining constraints similar to ours. Taken together, these features make a modified sound horizon an attractive solution that preserves the agreement between geometry and structure while addressing both the Hubble tension and DESI's dynamical dark energy signal.

In this scenario, current sound-horizon predictions bias geometry-only estimates of $h$ and $\Omegam$, with $\Omegam h^2$ fixed when assuming $w_0w_a$CDM. Adopting $\Omegam h^2 = 0.1405 \pm 0.0006$ from the geometry-only posterior, the SH0ES \citep{shoes25} and CCHP \citep{cchp25} values of $h$ imply $\Omegam=0.262\pm 0.005$ and $\Omegam = 0.28 \pm 0.01$ respectively, giving low-redshift structure probes a clear target for informing the current debate on the Hubble tension. The posteriors from Combos 1 and 2 in $\Lambda$CDM are $\Omegam=0.30\pm 0.01$ and $\Omegam=0.29\pm 0.01$ (see Fig.~\ref{fig:lowz}). Regardless of how these puzzles are resolved, our results provide a striking demonstration of how the complementarity between geometry and structure-growth probes can sharpen the case for new physics from cosmological data.

 \begin{table*}[]
    \centering
    \caption{Posteriors for the DES Y1 clusters and geometry-only data sets, and their combination, in flat $\Lambda$CDM and $w_0w_a$CDM. The parameter $r$ is the correlation coefficient between $\sigma_8$ and $\Omegam$. Cosmological parameters for DES Y1 clusters-only are omitted because they are limited by the emulator parameter range (see text). \vspace{5pt}}
    \begin{tabular}{ccccc}
    \hline 
      \hspace{0.05in} Model \hspace{0.05in} & \hspace{0.05in} Parameter \hspace{0.05in} & \hspace{0.05in} DES Y1 Clusters \hspace{0.05in} & \hspace{0.05in} Geometry-only \hspace{0.05in} & \hspace{0.05in} Geometry+Clusters \hspace{0.05in} \\
    \hline 
    $\Lambda$CDM & $\sigma_8$  & $0.826\pm0.032$ & $0.802\pm0.007$ & $0.797\pm0.006$ \\
    & $\Omega_m$ &  $0.254\pm0.023$ & $0.302\pm0.004$ & $0.299\pm0.004$ \\
    & $S_8$ & $0.759\pm0.019$& $0.804\pm0.009$ & $0.796\pm0.008$ \\
    & $r$ & $0.36$ & $-0.26$ & $-0.06$ \\  
    & $H_0$ $\mathrm{[km}$ $\mathrm{s}^{-1}$ $\mathrm{Mpc}^{-1}]$& $71.13\pm1.57$ &  $68.22\pm0.284$& $68.42\pm0.282$ \\ \hline \hline 
    $w_0w_a$CDM & $\sigma_8$  &  --- & $0.804\pm0.011$ & $0.802\pm0.010$\\
    & $\Omega_m$ & --- & $0.314\pm0.009$ & $0.293\pm0.007$\\
    & $S_8$ & --- & $0.823\pm0.012$ & $0.793\pm0.008$\\
    & $r$ & --- & 0.27 & 0.40\\
    & $H_0$ $\mathrm{[km}$ $\mathrm{s}^{-1}$ $\mathrm{Mpc}^{-1}]$  & --- & $67.17\pm0.96$ & $69.14\pm0.83$\\
    & $w_0$ & --- & $-0.810\pm0.101$ & $-1.038\pm0.066$\\
    & $w_a$ & --- & $-0.672\pm0.321$ & $0.048\pm0.188$\\
    & $D_w$ & --- & $0.698\pm0.335$ & $-0.058\pm0.198$\\
    \hline
    \end{tabular}
    \label{tab:constraints}
\end{table*}

\vspace{1em} \noindent \textbf{\emph{Acknowledgements.}} --- We thank David Weinberg for valuable discussions about this work, and comments on our manuscript which substantially improved the presentation of our work. We thank the \abacussummit\ team for making the simulation publicly available. ANS also thanks Cristhian Garcia-Quintero for his assistance with the {\sc{desilike}} code. ANS and ER received funding for this work from the Department of Energy (DOE) grant DE-SC0009913. ANS received funding for this work from DOE grants DE-SC0020247 and DE-SC0025993. HW and SC are supported by the DOE Award DE-SC0010129 and the NSF Award AST-2509910. This material is based upon work supported in part by the National Science Foundation through Cooperative Agreements AST-1258333 and AST-2241526 and Cooperative Support Agreements AST-1202910 and AST-2211468 managed by the Association of Universities for Research in Astronomy (AURA), and the Department of Energy under Contract No.\ DE-AC02-76SF00515 with the SLAC National Accelerator Laboratory managed by Stanford University. Additional Rubin Observatory funding comes from private donations, grants to universities, and in-kind support from LSST-DA Institutional Members. Simulations were analyzed in part on computational resources of the Ohio Supercomputer Center \cite{OhioSupercomputerCenter1987}, with resources supported in part by the Center for Cosmology and AstroParticle Physics at the Ohio State University. We gratefully acknowledge the use of the {\sc{matplotlib}} software package \cite{Hunter_2007} and the GNU Scientific library \cite{GSL_2009}. This research has made use of the SAO/NASA Astrophysics Data System.


\bibliographystyle{apsrev}
\bibliography{masterbib,rozobib}

\clearpage
\onecolumngrid
\appendix
\section{Supplemental Material}

\subsection{Choice of Experiments}

When selecting experiments for defining Combo-1 and Combo-2 our goal was to select experiments that were both mutually independent, and representative of the current state of the art.  We rely on the X-ray cluster constraints from \citet{Mantz_et_al_2015} rather than the more recent constraints from \textit{eRosita} because the latter reports $\gtrsim3\sigma$ tension with DES, KiDS, HSC, and SDSS clustering \citep{Ghirardini_et_al_2024}. We did not use the KiDS-1000 3$\times$2-point analysis due to the reported improvement in photo-$z$ accuracy in the more recent cosmic shear paper (item 8) above.  We also did not use the HSC 3$\times$2-point analysis because it relies on SDSS clustering data, resulting in non-trivial covariance with the full-shape analysis of SDSS galaxy clustering (item 3).  While there are multiple full-shape analyses in SDSS \citep[e.g.][]{Philcox_et_al_2022, Pellejero-Ibanez_et_al_2024}, the posteriors from item 3 roughly encompass the range of posteriors in the literature \citep[lower-right panel of Fig.~4 in][]{Salcedo_et_al_2025c}, which is why we selected it as our fiducial SDSS full-shape constraint.

The four experiments in each Combo are nearly independent of one another. If we compare experiments across the two Combos, the strongest overlap is that between the DES 3$\times$2-point and SPT cluster abundances in Combo-1, and the DES Y1 cluster abundances in Combo-2.  However, only $\sim1\%$ of the DES Y1 $\lambda > 10$ clusters used in the analysis of \citet{Salcedo_et_al_2025c} are shared with the SPT analysis of \citep{Bocquet_et_al_2024}.  Moreover, the DES Y1 cluster posteriors are shape noise dominated, while the DES Y3 3$\times$2 posteriors are sample variance dominated, with DES Y3 sharing only $\approx 20\%$ of its area with DES Y1. Consequently, Combos 1 and 2 are also nearly mutually independent.

\subsection{Combining Experiments}

Throughout this paper, unless stated otherwise, when we combine two data sets $A$ and $B$ we use the code {\sc CombineHarvesterFlow} \citep{Taylor_CombineHarvesterFlow_et_al_2024}.  This code fits a normalizing flow to the published MCMC chain of each experiment separately, and uses the resulting flow-based approximation to importance sample the chain from one experiment using the other.  Combining the cluster constraints with other experiments requires some care due to the fact that the cluster data always assumed a fixed $\theta_*$, a problem we now address. 

Consider two experiments $A$ and $B$, each of which constrains a series of parameters $x$, as well as a special parameter $y$ which is held fixed by one of the two experiments.  The posteriors for experiments $A$ and $B$ are $P_A(x,y)$ and $P_B(x|y=y_0)$.  We wish to determine $P_{A+B}(x,y)$, the joint posterior.  To do so, we assume that the parameter $y$ is tightly constrained by experiment $A$, but not so by experiment $B$. Assuming the two experiments $A$ and $B$ are independent one has
\begin{align}
    P_{A+B}(x,y) & = P_A(x,y)P_B(x,y) \\
    & = P_A(x|y)P_A(y)P_B(x|y)P_B(y).
\end{align}
If $P_A(y)$ is sharply peaked around $y_0$, whereas $B$ is largely insensitive to $y$, then one further has
\begin{align}
    P_{A+B}(x,y) & \propto P_A(x|y)P_A(y)P_B(x|y_0) \\
    & = P_A(x|y)P_B(x|y_0).
\end{align}
That is, one can importance sample experiment $A$ given a chain of experiment B at fixed $y_0$, by simply ``ignoring $y$'' when computing the importance sampling weight for chain $A$, but then importance sampling the full $A$ chain.

Combining clusters with 3$\times$2-point requires a different approach.  In this case, one simply doesn't know $P_B(y)$, so reconstructing $P_{A+B}(y)$ is impossible.  Instead, we simply apply a CMB-based $\theta_*$ prior to experiment $A$, reducing the problem to the case we treated before.  Note this means that when we combined 3$\times$2-point data with DES clusters, the resulting posterior always includes a CMB-based $\theta_*$ prior.  As an example, in the middle panel of Fig.~\ref{fig:lowz}, the combined constraint results in a tight constraint on $h$ due to the CMB-based $\theta_*$ prior used when combining the data sets.

\end{document}